\documentclass[fleqn,twoside]{article}
\usepackage{espcrc2}

\usepackage{graphicx}
\usepackage[figuresright]{rotating}

\newcommand{\AmS}{{\protect\the\textfont2
  A\kern-.1667em\lower.5ex\hbox{M}\kern-.125emS}}

\title{Recent results from systematic parameterizations of \\
Ginsparg-Wilson
fermions}

\author{Christof Gattringer\address[MCSD]{
Institut f\"ur Theoretische Physik, 
        Universit\"at Regensburg,  
        D-93040 Regensburg, Germany}%
        \thanks{The work reported here was done in the BGR
(Bern-Graz-Regensburg) collaboration. The author is supported by the 
Austrian Academy of Sciences, APART 654.}}
       
\begin{document}

\begin{abstract}
The "Fixed Point Dirac Operator" and "Chirally Improved Fermions" both use
large numbers of gauge paths and the full Dirac structure to approximate 
a solution of the Ginsparg-Wilson equation. After a brief review of the 
two approaches we present recent results for quenched QCD with pion masses 
down to 210 MeV. We discuss the limits and advantages of approximate 
parameterizations and outline future perspectives.
\vspace{1pc}
\end{abstract}

\maketitle

\section{Introduction}

The fundamental difficulties with the implementation of chiral
symmetry on the lattice were finally overcome with the rediscovery of 
the Ginsparg-Wilson equation \cite{GiWi82,Ha98} for the lattice Dirac 
operator $D$, 
\begin{equation}
D \, \gamma_5 \; + \; \gamma_5 \, D \;\; = \;\; 2 a \, D \, R \gamma_5
\, D \; . 
\label{giwi}
\end{equation}
On the right-hand side $R$ is a local operator commuting with
$\gamma_5$ and $a$ denotes the lattice spacing. 
This equation implies exact chiral symmetry on the lattice 
\cite{Lu98}. Results obtained with Dirac
operators obeying (\ref{giwi}) exactly or approximately now allow
to test QCD also in the chiral regime (see \cite{giusti} for a
recent review). The most widely used chiral Dirac operator is the
so-called overlap operator \cite{overlap}. The overlap formula
gives a simple explicit prescription how to construct a Ginsparg-Wilson
fermion (i.e.~a solution of Eq.~(\ref{giwi})) from any decent 
lattice Dirac operator. An approach related to the overlap formalism 
are domain wall fermions \cite{domainwall} where an auxiliary fifth
dimension is used to implement the chiral symmetry. 

Besides the overlap and domain wall fermions two more approaches to
chiral symmetry are known, fixed point fermions \cite{fp1,fp2,fp3,fp4}
and the chirally improved operator \cite{ci1,ci2}. The fixed point (FP)
Dirac operator is 
constructed from the saddle point evaluation of the RG equations. 
The chirally improved (CI) operator is obtained by a systematic
expansion of a solution of the Ginsparg-Wilson equation. In a
practical application both the FP and the CI operator will use only 
finitely many terms (essentially restricted to the hypercube) and one can 
expect \cite{diracsupport} only an approximation of a Ginsparg-Wilson
fermion. Exact chiral symmetry can, however, be obtained by using
the FP or CI operators as a starting point in the overlap projection
(see also \cite{impoverlap}).
Using an improved operator for the overlap was found 
to improve the localization and one can hope to also 
obtain better dispersion and scaling properties. However, additional
overlap steps also drive up the cost of the Dirac operator in
numerical implementations. 

\begin{figure}[ht]
\hspace{-3mm}
\includegraphics[width=8cm]{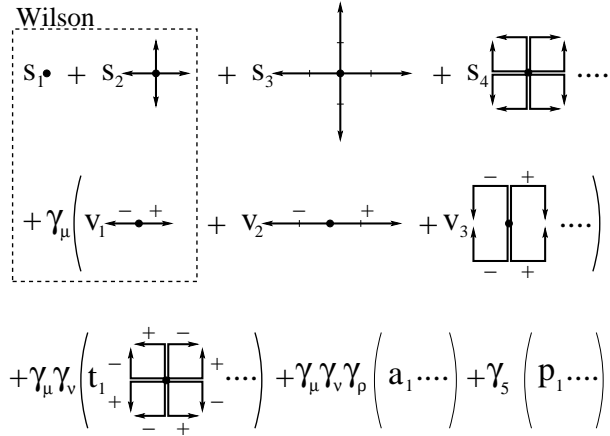}
\vspace{-7mm}
\caption{Schematic representation of a general lattice Dirac operator.}
\label{diracschem}
\vspace{-3mm}
\end{figure}

In this contribution we analyze how far into the chiral region one can 
proceed with FP or CI fermions without additional overlap steps. 
Following a short review of the construction of FP and CI fermions
we will discuss several recently obtained results. We
address the spectrum of the Dirac operator and the determination of
the topological charge and susceptibility from the index theorem. This
is followed by a discussion of the results from quenched spectroscopy 
with emphasis on the pion mass close to the chiral limit where we extract
the quenched chiral log parameter $\delta$. We study the scaling
behavior of rho and proton masses and of the pion decay constant. 
We conclude with a discussion of the range of pion masses 
where one can work competitively with FP and CI fermions without 
overlap projection.
 
\section{Construction of FP and CI fermions}

The construction for both the FP and CI operators starts from 
the expression for a general Dirac operator on the lattice
\cite{fp2,ci1}. In Fig.~\ref{diracschem} we give a schematic representation of 
such a general lattice Dirac operator.

Let us start the discussion of the structure with 
already familiar terms. In the dashed square in the upper 
left corner of Fig.~\ref{diracschem}
we show the terms that are used in the Wilson Dirac operator. 
There are two terms coming with the unit matrix in Dirac space: 
Firstly a
constant term consisting of the mass parameter and the center term in
the discretization of the Laplacian. They are represented by a
dot. Secondly also the nearest neighbor terms from the discretization
of the Laplacian come with a unit matrix in Dirac space. These hopping
terms are represented by single hops, i.e.~straight lines in all
directions (for artistic reasons we show only two of the four possible 
directions). Besides the two terms trivial in Dirac space, the Wilson 
operator also contains the naively discretized massless Dirac operator
where a sum over all $\gamma_\mu$ together with hops in positive and 
negative $\mu$-direction occurs (we show only one of the four
directions). In order to describe a derivative here the hops in 
positive and negative direction come with different signs. The
parameters $s_1,s_2$ and $v_1$ are real numbers chosen such that 
a fermion free of doublers with a given mass is described.

A more general Dirac operator is obtained when more terms are
used to discretize the derivative, such as next to nearest neighbor terms,
staples etc. Similarly one can also extend the number of terms
in the trivial Dirac sector. Finally from the Symanzik improvement
program it is already known that also sectors of the Clifford algebra 
other than the trivial and vector sectors can contribute to the Dirac
operator. So the structure of a general Dirac operator as depicted in 
Fig.~\ref{diracschem} is a sum over all elements of the Clifford
algebra. Each element is multiplied with paths of link variables and
each path has some coefficient $s_i, v_i, t_i, a_i, p_i$. Some of 
the paths have the same coefficient but differ by relative sign
factors. These sign factors are entirely determined by implementing
the symmetries C,P and $\gamma_5$-hermiticity (i.e.~$\gamma_5 D
\gamma_5 = D^\dagger$). Rotation invariance requires paths related by
rotations on the lattice to come with the same coefficient and
translation invariance makes the coefficients independent of the
actual space time point. With our particular choice of the
representation of the Clifford algebra the coefficients $s_i, v_i, t_i,
a_i, p_i$ are real. A further generalization can be obtained by
allowing these coefficients to be real functions of local loops of
gauge links. This option is used for the construction of the FP Dirac 
operator. 

The FP and the CI operator now differ only in the method for
determining the coefficients $s_i, v_i, t_i, a_i, p_i$. For the FP
operator the goal is to use the coefficients to approximate a
solution of the fixed point equation for the Dirac operator. The 
basic idea is that near the critical surface the theory is invariant
under real space renormalization group transformations that relate
the quantum fields on a fine lattice to fields on a coarser lattice.
In general the corresponding equations are quite involved but in the
weak coupling limit they can be solved using the saddle point
method. For the Dirac operator the FP equation reads
\begin{equation}
D_c \;  \; = \; \; \kappa \; - \; \kappa^2 \, \Omega \, \left[ \, D_f \,
+ \ \kappa \Omega^\dagger \Omega \, \right]^{-1}\Omega^\dagger \; .
\label{fpequation}
\end{equation}
Here $D_c$ ($D_f$) is the FP Dirac operator evaluated on the coarse
(fine) gauge field configurations.
$\Omega$ is the blocking kernel for the fermions and
$\kappa$ a free parameter of the blocking procedure which can be used 
to maximize the slope of the exponential decay of the Dirac operator.
The coefficients $s_i, v_i, t_i,
a_i, p_i$ of the Dirac operator were now adjusted such that a 
$\chi^2$ functional measuring essentially the difference 
between the left-hand and
right-hand sides of (\ref{fpequation}) on an ensemble of coarse and
fine gauge configurations related by the renormalization group
is minimized. In the actual
construction this procedure was iterated starting from an ensemble at
very weak coupling, followed by an intermediate step and finally a last
minimization step at the target coupling. We remark that the action
was optimized only for one such coupling corresponding to a lattice
spacing of $0.16$ fm and then used also for the other two
couplings. The parameterized FP operator is described by 82 couplings
corresponding to 41 independent terms and each of the coefficients $s_i, v_i, t_i,
a_i, p_i$ is a linear function of local closed gauge loops. All
terms of the FP operator are restricted to the hypercube. 

For the CI operator the strategy is to directly insert the expanded
Dirac operator as depicted in Fig.~\ref{diracschem} into the
Ginsparg-Wilson equation (\ref{giwi}) with a trivial kernel $R = 1/2$.
The evaluation of the two sides of Eq.~(\ref{giwi}) 
can be implemented in an algebraic computer program using
directly the systematic representation of Fig.~\ref{diracschem}.
The result is an expansion for both sides of the Ginsparg-Wilson
equation similar to the expansion of the original Dirac operator. One
can show that the individual terms (different element of the Clifford
algebra, different paths) are linearly independent. The coefficients
of equal terms on the two sides have to be equal and one can read off a
system of coupled quadratic equations for the expansion coefficients
\begin{eqnarray}
2 s_1 & = & s_1^2 + 8s_2^2 ... + 8v_1^2 ... \nonumber \\
2 s_2 & = & 2 s_1 s_2 + 12 s_2 s_3 ... + 12 v_1 v_2 ... \nonumber \\
..... & & 
\label{ciequation}
\end{eqnarray}
This system is equivalent to the original Ginsparg-Wilson
equation. After one truncates the expansion in Fig.~\ref{diracschem}
this system becomes finite and can be solved numerically. In addition 
to the equations representing the Ginsparg-Wilson equation 
one can impose boundary conditions, i.e.~additional equations 
for the coefficients enforcing vanishing quark mass, the
correct dispersion relation in the free case, ${\cal O}(a)$ improvement etc.
The resulting CI operator is an approximation of a solution of the
Ginsparg-Wilson equation. The CI operator has 19 coefficients and
terms on the hypercube plus an extra L-shaped term of length $\sqrt{5}$.  

\section{Spectrum of eigenvalues}

A first impression of the quality of the approximation of a
Ginsparg-Wilson Dirac operator can be obtained by inspecting eigenvalues
of the Dirac operator in typical gauge backgrounds. For an exact
solution of the Ginsparg-Wilson equation the spectrum is restricted
to a circle of radius 1 with center 1 in the complex plane. For the
two approximations considered here the spectrum will not exactly lie
on the Ginsparg-Wilson circle but show small fluctuations around it. 

\begin{figure}[t]
\hspace*{18mm}
\includegraphics[height=6cm]{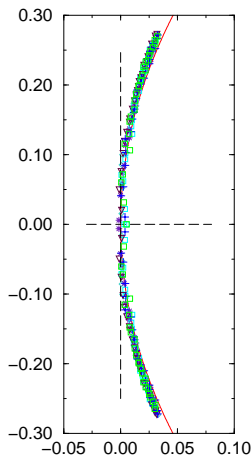}
\vspace{-7mm}
\caption{Superposition of 6 spectra of the CI operator. Shown are the 50
smallest eigenvalues in the complex plane for 6 configurations on 
$12^3 \times 24$ lattices with lattice spacing $a = 0.1$ fm.}
\label{fig:typspect}
\vspace{-3mm}
\end{figure}

In Fig.~\ref{fig:typspect} we show a superposition of 6 spectra of the
CI operator in quenched background gauge configurations on 
$12^3 \times 24$ 
lattices with lattice spacing $a = 0.1$ fm. Shown are the 50 smallest
eigenvalues for each configuration. It is obvious that the eigenvalues
are not located exactly on the Ginsparg-Wilson circle but the
fluctuations are rather small. In particular we do not find many
configurations with negative real eigenvalues. Such so-called exceptional
configurations give rise to numerical problems in the computation of
the propagator and limit the smallest quark masses one can work at. As
will be discussed below the suppression of the eigenvalue fluctuations 
achieved by the FP and CI operators allows us to work at considerably
smaller quark masses than with e.g.~Wilson fermions.

An immediate consequence of the well ordered spectrum is the
possibility to use the index theorem \cite{indexlat}
for evaluating the topological
charge with the FP and CI operators, while for Wilson fermions the
large fluctuations of the eigenvalues lead to a mixture of physical
and doubler modes \cite{wilsonfluc}. We determined the topological
charge as $Q = n_- - \, n_+$ where $n_-$ ($n_+$) is the number of
eigenvectors with negative (positive) matrix element with $\gamma_5$. 
Measurements \cite{fp3,GaHoSch02} on different volumes $V$ and lattice 
spacings give 
for the topological susceptibility ($\chi_{top} = \langle Q^2 \rangle/V$)   
values of (196(4)MeV)$^4$ for the FP operator and (191(2)MeV)$^4$ for
the CI operator. 

\begin{figure}[h]
\hspace*{-1mm}
\includegraphics[height=6.4cm,clip]{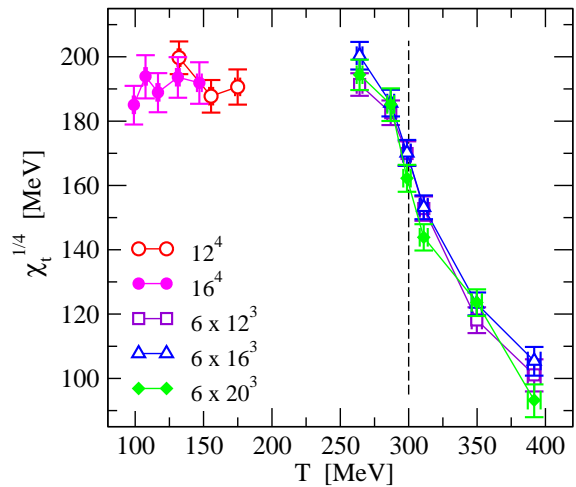}
\vspace{-7mm}
\caption{The behavior of the topological susceptibility across the
QCD phase transition computed using the eigenvalues of the CI operator 
and the index theorem. The critical temperature is marked by a dashed
vertical line.}
\label{fig:chivstemp}
\vspace{-3mm}
\end{figure}

For the CI operator a detailed analysis of the behavior of
$\chi_{top}$ across the finite temperature QCD phase transition was performed 
\cite{GaHoSch02}. The quenched gauge configurations were generated with
the L\"uscher-Weisz gauge action \cite{luweact}. Our results for the
topological susceptibility as a function of the temperature are
displayed in Fig.~\ref{fig:chivstemp}. One finds that the 
results obtained on $6 \times L^3$ lattices with spatial extent $L =
12,16$ and 20 nicely agree with each other showing that finite size 
effects are under control. We also include results from $12^4$ and
$16^4$ lattices to set the base line below $T_c$. The critical
temperature $T_c$ as determined for the L\"uscher-Weisz action in 
\cite{GaRaSchSo02} is marked by the dashed vertical line. 
One finds that the topological susceptibility starts to decrease
already below $T_c$ and quickly diminishes as the temperature
increases further. The results obtained here using the index theorem
agree well with calculations based on bosonic methods for the
determination of the topological charge \cite{AlElDG97}.

\section{Quenched spectroscopy}

During the last year the major goal of the BGR collaboration was to
use the FP and CI Dirac operators for quenched spectroscopy. Before we
discuss the results of this study let us briefly outline the setting 
of these calculations. 

For the FP operator the perfect gauge action \cite{perfgauge} was
used to generate the quenched ensembles. The gauge
configurations were smoothened with perfect smearing \cite{fp3} (which
we consider as part of the parameterization of the Dirac operator) and
subsequently fixed to Coulomb gauge. 
For the quark sources a Gaussian distribution was used. 

The CI operator was used with gauge ensembles generated with the 
L\"uscher-Weisz action \cite{luweact}. Here the gauge configurations
were treated with one step of hypercubic blocking \cite{hypblocking}.  
For the CI operator Jacobi smeared sources \cite{jacobi} were used and 
no gauge fixing was necessary. 

For both operators we worked on three lattice sizes, $8^3 \times 24, 12^3
\times 24$ and $16^3 \times 32$. For both gauge actions we used 
three different gauge
couplings corresponding to lattice spacings of roughly $a = 0.15$
(0.16 for FP), 0.10 
and 0.08 fm ($a$ determined from the Sommer parameter). The statistics 
varied between 100 and 200 configurations for each ensemble. Our
choice for the volumes and lattice spacings allows to study the
scaling behavior with three different values of $a$ at a fixed
spatial length of about 1.3 fm and a finite size study at a fixed lattice
spacing $a = 0.16$ fm. For a more detailed account of our setting 
for quenched spectroscopy see \cite{bgrlat02}.

\begin{figure}[t]
\hspace*{-2mm}
\includegraphics[height=6.5cm,clip]{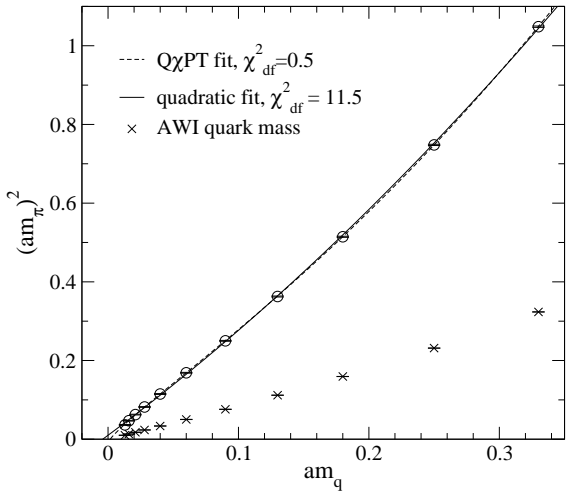}
\vspace{-7mm}
\caption{Pion mass (circles) and AWI mass (crosses) as a function of
the bare quark mass. $16^3 \times 32$
lattice, $a = 0.16$ fm, FP operator.}
\label{fig:mpivsmquark}
\vspace{-1mm}
\end{figure}

Let us begin the discussion of our quenched spectroscopy results with
the pion mass and the axial Ward identity (AWI) mass. We computed the
pion mass using different 2-point correlators (pseudoscalar, 
time component of the axial current, and the mixed correlator) and
also compared correlators with smeared sink to point-like sink
correlators. For all these measurements we find good agreement of the pion
masses. The (unrenormalized) AWI mass $m_{AWI}$ was computed as
\begin{equation}
m_{AWI} \; = \; \frac{ \langle \partial_0 A_0 P \rangle}{ 2 \, \langle P
P \rangle } \; ,
\end{equation}
where $P$ is the pseudoscalar density and $A_0$ the time component of
the axial vector current. 
In Fig.~\ref{fig:mpivsmquark} we show the mass of the
pion as computed from the pseudoscalar 2-point function (circles) 
and the AWI mass (crosses) as a 
function of the bare quark mass. This plot was generated using the
FP operator on a $16^3 \times 32$ lattice at a lattice spacing $a =
0.16$ fm.  

It is a convincing sign of good chiral properties that both the pion
mass and the AWI mass extrapolate to zero 
with the bare quark mass. However, for smaller 
volumes ( $L = 1.8$ or 1.3 fm) we found that the quenched topological
finite size effects caused by the zero modes \cite{pioneffects} become
more important.
They can be removed by subtracting the scalar propagator, which has the
same topological finite size effect but larger mass, from the
pseudoscalar 2-point function (see \cite{bgrlat02} for more details). 
In Fig.~\ref{fig:mpivsmquark} we show also two fits to the pion data,
a quadratic fit (full curve) and a fit including the quenched chiral
log \cite{quenchlog} (dashed curve). Our results for the quenched
chiral log from a different method will be discussed below. 

Another test of chirality is to determine the residual quark
mass. This was done using a linear fit to the AWI mass and taking
the intercept of the straight line with the horizontal axis as the
residual quark mass $m_{res}$. For the FP operator we computed a
residual mass of $a \, m_{res} = -0.0006(4)$ at $a = 0.16$ fm increasing
in size to $a \, m_{res} = -0.0194(2)$ at $a = 0.08$. This increase is
due to the fact that we used the FP action which was optimized for 
the $a = 0.16$ ensemble also at finer lattice spacing without
redetermining the coefficients. For the CI operator the residual quark
mass came out between $a \, m_{res} = 0.002(1)$ at $a = 0.15$ fm
and  $a \, m_{res} = 0.000(1)$ at $a = 0.08$ fm. The smallest pion masses
we have worked at are $m_\pi = 210$ MeV for the FP operator and 
$m_\pi = 240$ MeV for the CI operator. We expect that for both
operators it is possible to go down to $m_\pi \sim  200$ MeV without
having to use exceedingly fine lattices.

\begin{figure}[t]
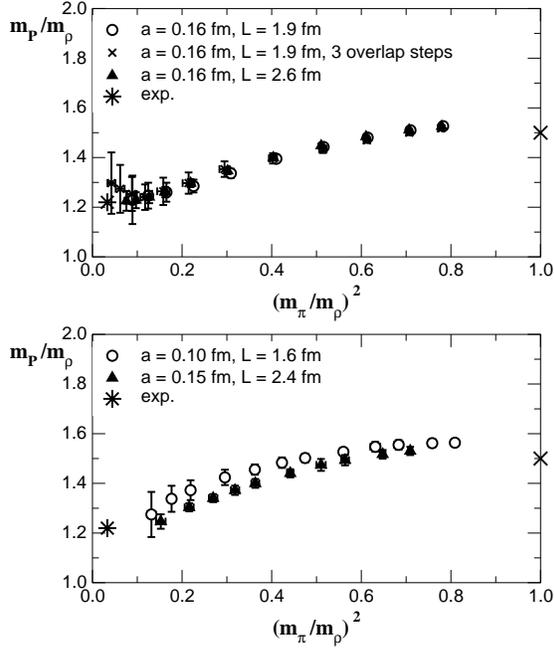

\includegraphics[height=4.3cm,clip]{ape_fp.eps} \\
\includegraphics[height=4.3cm,clip]{ape_ci.eps}
\vspace{-7mm}
\caption{APE plots for the FP (top) and CI (bottom) operators.}
\label{fig:apeplots}
\vspace{-3mm}
\end{figure}

As another important benchmark measurement we looked at APE and
Edinburgh plots. In Fig.~\ref{fig:apeplots} we show APE plots 
for the FP (top) and the CI (bottom) operator. For the FP operator 
we show three sets of data: The results for spatial extent 
$L = 1.9$ fm at $a = 0.16$ fm (circles), 
the same ensemble but with the FP operator augmented with three
steps of the overlap projection using Legendre polynomials (crosses) 
and finally results at $L = 2.6$ fm, $a = 0.16$ fm (triangles).
When comparing the FP operator on volumes with $L = 1.9$ fm and $L = 2.6$ fm
we find that the data agree well and we do not observe finite
size effects for $L \ge 1.9$ fm. Also the overlap augmented FP operator
(crosses) gives rise to results which are in very good agreement with
the unimproved FP data. 

In the bottom figure for the CI operator we compare a data set from
a finer but smaller lattice ($a = 0.1$ fm, $L = 1.6$ fm, 
circles) to results from a lattice similar to the one used for the FP 
operator in the top figure ($a = 0.15$ fm, $L = 2.4$ fm, triangles).   
Here we do see a splitting between the two curves which we attribute
to finite size effects and a small scaling violation. It is
interesting to note that the data from the FP operator extrapolate
very well to the physical value (marked by a star in the plot), while
the result from the CI operator undershoots the physical value similar
to what is known from other quenched simulations (see
e.g.~\cite{cppacsquench}).

\begin{figure}[t]
\hspace*{2mm}
\includegraphics[width=6.5cm,clip]{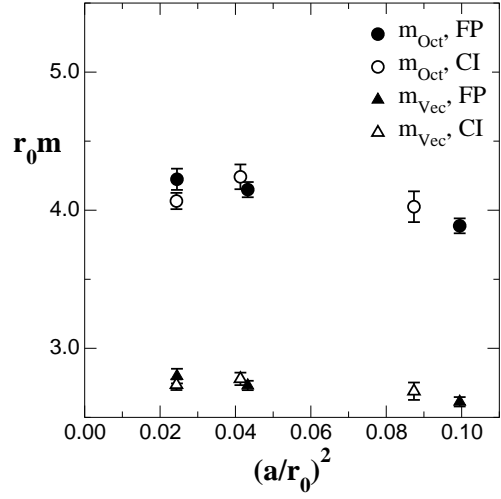}
\vspace{-7mm}
\caption{Scaling of the octet baryon mass (circles) and the vector 
mass (triangles). Filled symbols represent the FP results, while open symbols
are used for the CI operator.}
\label{fig:masscaling}
\vspace{-3mm}
\end{figure}

We conclude our discussion of the quenched spectroscopy with a brief
discussion of the scaling properties of hadron masses. To do so, we
work for both the FP and the CI operator at a bare quark mass which gives  
a ratio of $m_\pi/m_\rho = 0.7$. At such a mass the statistical error
is small and also a comparison with less chiral actions can be done
(see e.g.~\cite{fp2}). In Fig.~\ref{fig:masscaling} we show the octet
(circles) and vector (triangles) masses as a function of the 
square of the lattice spacing.
For both the horizontal and vertical axes we use the Sommer parameter
to set the scale. Filled symbols represent the FP results, while open symbols
are used for the CI operator. The symbols are connected to guide the eye.

The results from the two operators agree within error bars. Both sets
of data show only a small deviation from a horizontal line indicating
that both ${\cal O}(a)$ and ${\cal O}(a^2)$ effects are small. Note
that since both the parameterized FP operator as well as the CI
operator are only approximate solutions of the Ginsparg-Wilson it can
not a priori be expected that ${\cal O}(a)$ effects are absent. However, the
plot shows that scaling violations are very small. 

\section{Pion decay constant}

The definition of the pion decay constant can be combined with the
axial Ward identity to yield
\begin{equation}
f_\pi \; = \; \frac{ 2m \, \sqrt{Z_{PP} }}{m_\pi^2}.
\label{fpi}
\end{equation}
Here $Z_{PP}$ is the prefactor of the pseudoscalar 2-point function
at zero-momentum, 
\begin{equation}
\sum_x \langle P(x,t) P(0,0) \rangle \; \sim \; \frac{Z_{PP}}{2 m_\pi} \,
\exp(\, - \, m_\pi t \,) .
\end{equation}
Strictly speaking formula (\ref{fpi}) holds only for Dirac operators
with exact chiral symmetry where the product of renormalization
constants $Z_m Z_P$ equals 1. For the approximate solutions of the 
Ginsparg-Wilson equation which we use here this is currently only an
assumption which will eventually have to be tested.

\begin{figure}[t]
\hspace*{-2mm}
\includegraphics[height=48mm,clip]{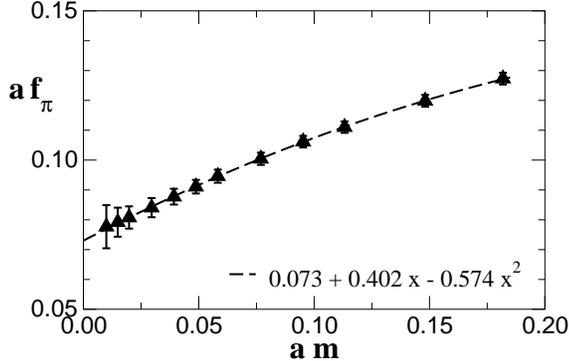}
\vspace{-7mm}
\caption{Pion decay constant as a function of the bare quark mass. 
$16^3 \times 32$
lattice, $a = 0.1$ fm, CI operator.}
\label{fig:fpivsmquark}
\vspace{-2mm}
\end{figure}

In Fig.~\ref{fig:fpivsmquark} we show $f_\pi$ according to
Eq.~(\ref{fpi}) as a function of the
bare quark mass (both in units of the lattice spacing as determined
from the Sommer parameter) for the CI operator on the $16^3\times 32$ 
ensemble at $a = 0.1$ fm. We interpolate the data using a second order
polynomial which we subsequently use to extrapolate the data to the chiral 
limit. We collected the results from the chirally
extrapolated values of $f_\pi$ from the CI operator and analyzed the
scaling of this observable.  
We find that all data agree well
within error bars and the discretization errors are small. Our 
quenched results come out 10-15 \% larger than the experimental value
(see \cite{bgrlat02} for a more detailed discussion). 

\section{Quenched chiral logs}

We have already briefly addressed the possibility of extracting the
quenched chiral log parameter $\delta$ from the pion mass 
(compare Fig.~\ref{fig:mpivsmquark} and the discussion
below the figure). However, here we use a method \cite{cppacsquench} 
which allows to increase the
statistics by using pions with non-degenerate quark masses and which
also gets rid of the dependence on the unknown chiral perturbation 
scale $\Lambda_\chi$.

The idea is to form the combinations $x$ and $y$
\begin{eqnarray}
x &=& 2 \, + \, \frac{m_1 + m_2}{m_1 -m_2} \ln\left( \frac{m_2}{m_1}
\right) \; ,
\nonumber \\
y &=& \frac{4 \, m_1 m_2 }{(m_1 + m_2)^2} \, \frac{{M_{12}}^4}{{M_{11}}^2
{M_{22}}^2} \; .
\label{xyformula}
\end{eqnarray}
Here $M_{12}$ denotes the mass of the pseudoscalar with quark masses
$m_1$ and $m_2$. Using the results of \cite{quenchlog} one finds
\begin{equation}
y \; = \; 1 \; + \; \delta \, x \; + \;  {\cal O}(m^2, \delta^2, a_\phi) \;,
\label{deltaformula}
\end{equation}
i.e.~to leading order the quenched chiral log parameter can be read
off from the slope in a plot of $y$ versus $x$. 

\begin{figure}[t]
\hspace*{-3mm}
\includegraphics[width=7.7cm,clip]{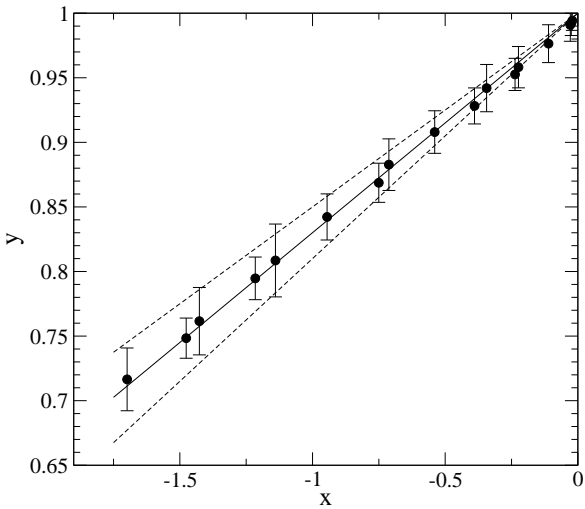}
\vspace{-7mm}
\caption{$x$-$y$ plot used to determine the quenched chiral log parameter
$\delta$ (FP operator, $16^3 \times 32, a = 0.16$ fm).}
\label{fig:xyplot}
\end{figure}

In Fig.~\ref{fig:xyplot} we show such a $x$-$y$ plot for the FP operator 
on the $16^3 \times 32$ lattice at $a = 0.16$ fm. The data lie inside a band
which gives rise to $\delta = 0.17(2)$ (compare \cite{quenchlogresults}
for previous determinations of $\delta$ with chiral fermions). 
For the CI operator we obtain a very similar result of $\delta =
0.18(3)$. These results were obtained by using the unrenormalized 
AWI quark mass for $m_1,m_2$ in Eq.~(\ref{xyformula}). For the CI
operator we have also experimented with using the bare quark mass
instead and we present a more detailed discussion in \cite{bgrlat02}.

We remark that we also have obtained preliminary results for the pion
scattering lengths and some details are also presented in \cite{bgrlat02}.

\section{Conclusions}

In this contribution we have presented results from quenched QCD
calculations using the parameterized FP and CI operators. Both these 
operators are approximate solutions of the Ginsparg-Wilson equation
and are expected to have good chiral properties. Here these expectations
are tested using various observables.  

We find that both the pion and the AWI mass nicely extrapolate to 0 as a
function of the bare quark mass. If one uses the offset in a linear 
extrapolation of the AWI mass as a criterion for the remaining chiral 
symmetry breaking we find residual quark masses between 1 and 4 MeV on 
a quite coarse lattice with $a = 0.16$ fm. We have successfully
extracted the quenched chiral log parameter $\delta$ using pions with
non-degenerate quark masses. The octet and vector masses, as well as
the pion decay constant show very good scaling behavior.  We have
demonstrated that we can reach pion masses of about 210 MeV without
having to go to very small lattice spacings. A more detailed account
of our measurements can be found in \cite{bgrlat02}.

\begin{figure}[t]
\hspace*{-3mm}
\includegraphics[width=7.5cm,clip]{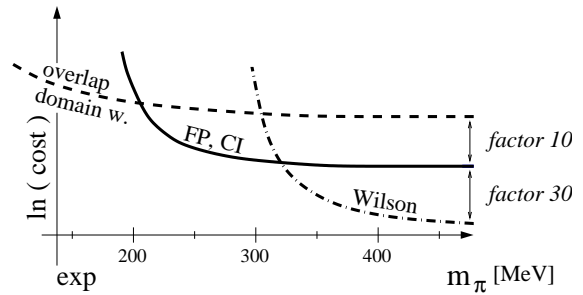}
\vspace{-3mm}
\caption{A sketch comparing the cost of different Dirac operators as a
function of the pion mass.}
\label{fig:cost}
\end{figure}

We would like to wrap up our conclusions with a brief comment on where
we think our approximate Ginsparg-Wilson fermions will be
competitive. In Fig.~\ref{fig:cost} we show a sketch which compares
the cost of different lattice Dirac operators as a function of the
pion mass. For heavy quarks the operator of choice is certainly
Wilson's Dirac operator, maybe combined with clover
improvement. However, if one wants to go to smaller masses one has to
increase the inverse gauge coupling to reduce eigenvalue fluctuations
responsible for exceptional configurations. This leads to a finer lattice 
and the volume has to be increased which drives up the cost, making
Wilson fermions very expensive below pion masses of 300 MeV. At such
small masses it is more economical to use the more expensive FP or CI
operators which cost about a factor of 30 more than Wilson's
operator (see \cite{bgrlat02} for a more detailed discussion of the
cost). As already remarked, with these operators we reached 210
MeV without having to go to very fine lattices and we expect that we
will remain competitive down to at least 200 MeV. If one needs to go
further into the chiral region, e.g.~when computing the chiral
condensate, one eventually has to use overlap projection to further
decrease the pion mass. We believe that in the window between pion
masses of 200 and 300 MeV the FP and CI operators are the best choice
and we will explore whether this is sufficient to make reliable contact   
with chiral perturbation theory.   

\newpage
\noindent
{\bf Acknowledgements:} I would like to thank the members of the BGR
collaboration for sharing their experience and enthusiasm during this
first year of work. The calculations were done on the Hitachi SR8000
at the Leibniz Rechenzentrum in Munich and we thank the LRZ staff for
training and support. This work was supported in parts by DFG and BMBF.


\begin{thebibliography}{9}

\bibitem{GiWi82} P.~Ginsparg, K.G.~Wilson, 
Phys.\ Rev.\ D 25 (1982) 2649.

\bibitem{Ha98}
P.\ Hasenfratz,
Nucl.\ Phys.\ Proc.\ Suppl.\ 63 (1998) 53.


\bibitem{Lu98}
M.\ L\"uscher,
Phys.\ Lett.\ B 428 (1998) 342.

\bibitem{giusti}
L.~Giusti, {\sl Exact chiral symmetry on the lattice: QCD
applications}, plenary talk at Lattice' 02, these proceedings.

\bibitem{overlap} 
R.\ Narayanan and H.\ Neuberger, Phys.\ Lett.\ B 302 (1993) 62,
Nucl.\ Phys.\ B 443 (1995) 305.

\bibitem{domainwall} 
D.B.\ Kaplan, Phys.\ Lett.\ B 288 (1992) 342;
Y.\ Shamir, Nucl.\ Phys.\ B 406 (1993) 90;
V.\ Furman and Y.\ Shamir, Nucl.\ Phys.\ B 439 (1995) 54.

\bibitem{fp1}
P.\ Hasenfratz, F.\ Niedermayer, Nucl.\ Phys.\ B 414 (1994) 785.

\bibitem{fp2} 
P.\ Hasenfratz, S.\ Hauswirth, K.\ Holland, T.\ J\"org, F.\
Niedermayer and U.\ Wenger,
Int.\ J.\ Mod.\ Phys.\ C 12 (2001) 691.

\bibitem{fp3} 
S.\ Hauswirth, {\sl Light hadron spectroscopy in quenched lattice 
QCD with chiral fixed-point fermions}, Thesis, Bern University 2002,
hep-lat/0204015;
T.\ J\"org, {\sl Chiral measurements in quenched lattice QCD with 
fixed-point fermions}, Thesis, Bern University 2002,
hep-lat/0206025.

\bibitem{fp4}
P.\ Hasenfratz, S.\ Hauswirth, T.\ J\"org, F.\ Niedermayer and K.\ Holland,
hep-lat/0205010.

\bibitem{ci1}
C.~Gattringer, Phys.~Rev.~D 63 (2001) 114501.

\bibitem{ci2}
C.\ Gattringer, I.\ Hip, C.B.\ Lang, Nucl.\ Phys.\ B 597 (2001) 451.

\bibitem{diracsupport}
I.~Horv\'ath,
Phys.\ Rev.\ Lett.\ 81 (1998) 4063,
Phys.\ Rev.\ D 60 (1999) 034510;
W.~Bietenholz,
hep-lat/9901005.

\bibitem{impoverlap}
W.~Bietenholz and I.~Hip,
Nucl.\ Phys.\ B 570 (2000) 423;
T.~DeGrand,
Phys.\ Rev.\ D 63 (2001) 034503;
W.~Bietenholz,
hep-lat/0204016.

\bibitem{indexlat}
P.\ Hasenfratz, V.\ Laliena and F.\ Niedermayer,
Phys.\ Lett.\ B 427 (1998) 125.

\bibitem{wilsonfluc}
C.~Gattringer and I.~Hip,
Nucl.\ Phys.\ B 536 (1998) 363,
Nucl.\ Phys.\ Proc.\ Suppl.\ 73 (1999) 871.

\bibitem{GaHoSch02}
C.~Gattringer, R.~Hoffmann and S.~Schaefer,
Phys.\ Lett.\ B 535 (2002) 358.

\bibitem{luweact}
M.\ L{\"u}scher and P.\ Weisz, Commun.\ Math.\ Phys.\ 97 (1985) 59;
Err.: 98 (1985) 433;
G.\ Curci, P.\ Menotti and G.\ Paffuti, Phys.\ Lett.\ B 130 (1983) 205,
Err.: B 135 (1984) 516.

\bibitem{GaRaSchSo02}
C.\ Gattringer, P.E.L.\ Rakow, A.\ Sch\"afer and W.\ S\"oldner,
Phys.\ Rev.~D (in print), hep-lat/0202009.

\bibitem{AlElDG97}
B.\ Alles, M.\ D'Elia and A.\ Di Giacomo, 
Nucl.\ Phys.\ B 494 (1997) 281.

\bibitem{perfgauge}
F.~Niedermayer, P.~R\"ufenacht and U.~Wenger, NPB, 2001.
Nucl.\ Phys.\ B 597 (2001) 413.

\bibitem{hypblocking} 
A.~Hasenfratz and F.~Knechtli,
Phys.~Rev.~D 64 (2001) 034504.

\bibitem{jacobi}
C.\ Best, M.\ G\"ockeler, R.\ Horsley, E.M.\ Ilgenfritz, H.\ Perlt,
P.E.L.\ Rakow, A.\ Sch\"afer, G.\ Schierholz, A.\ Schiller and S.\ Schramm
Phys.\ Rev.\ D 56 (1997) 2743;
C.\ R.\ Allton et al. (UKQCD Collaboration),
Phys.\ Rev.\ D 47 (1993) 5128.

\bibitem{bgrlat02}
C.~Gattringer et al. (BGR collaboration), these proceedings,
hep-lat/0209099.

\bibitem{pioneffects}
T.~Blum {\it et al.},
hep-lat/0007038;
S.J.\ Dong, T.\ Draper, I.\ Horv\'ath, F.X.\ Lee, K.F.\ Liu and J.B.\ Zhang,
Phys.\ Rev.\ D 65 (2002) 054507.

\bibitem{quenchlog}
S.R.~Sharpe, Phys.~Rev.~D 46 (1992) 3146;
C.W.\ Bernard and M.F.L.\ Golterman, Phys.\ Rev.\ D 46 (1992) 853.   

\bibitem{cppacsquench}
S.~Aoki et al. (CP-PACS collaboration), 
hep-lat/0206009.

\bibitem{quenchlogresults}
T.~W.~Chiu and T.~H.~Hsieh,
Phys.\ Rev.\ D 66 (2002) 014506;
T.\ Draper, S.J.\ Dong, I.\ Horv\'ath, F.X.\ Lee, K.F.\ Liu, N.\
Mathur and J.B.\ Zhang, hep-lat/0208045.

\end{thebibliography}
\end{document}